\documentstyle[preprint,aps,epsf]{revtex}
\tightenlines
\draft
\begin{document}

\title{Non-iterative and exact method for constraining particles in
a linear geometry}
\author{Horacio Tapia-McClung and Niels Gr{\o}nbech-Jensen}
\address{Department of Applied Science \\
University of California, Davis, California 95616}
\date{September 16, 2004}
\maketitle
\begin{abstract}
We present a practical numerical method for evaluating the
Lagrange multipliers necessary for maintaining a constrained linear geometry
of particles in dynamical simulations. The method involves no iterations,
and is limited in accuracy only by the numerical methods for solving
small systems of linear equations. As a result of the non-iterative and
exact (within numerical accuracy) nature of the procedure there is
no drift in the
constrained geometry, and the method is therefore readily applied
to molecular dynamics simulations of, e.g., rigid linear molecules or
materials of non-spherical grains. We illustrate the approach through
implementation in the commonly used second-order velocity explicit
Verlet method.
\end{abstract}

\vskip2pc
\narrowtext
\newpage
\section{Introduction}
Computational studies of the dynamics and statistics of large particle
ensembles have been successfully conducted for several decades
\cite{overview_1,overview_2}, and many different types of inter-particle force
fields have been developed for specific physical applications. The particles
of interest in Molecular Dynamics (MD) or Discrete Element Methods (DEMs) are
often represented by spatial point coordinates and the associated
momenta. For example, atomic-scale modeling of materials considers each
atom as a point particle, and the nature of the material is defined by the
interatomic potential. Likewise, the dynamics of inhomogeneous atomic
ensembles, such as molecular chains in solution, can be modeled by
point particles with different interaction potentials defining the nature
of the chemistry between two or more atoms. Given a complex ensemble of
interactions, the involved time scales of the dynamics may span many orders
of magnitude, thereby limiting the efficiency of a simulation approach since
a useful numerical time step for a simulation is inversely proportional to
the highest frequency in the system. Thus, if particles interact such that
high-frequency, small-amplitude oscillations result, it may be advantageous
to disregard the dynamics of the interaction and constrain the relative
degree of freedom between the two objects to the mean distance. In molecular
systems at room temperature conditions, constraints are further justified
by the fact that many relevant chemical bond vibrations are improbable to
be found in an excited quantum mechanical energy state. Other examples
of the application of holonomic constraints can be found in studies
of granular materials composed of non-spherical particles. Many such
studies relate
to particles of complex geometry or to particles that are connected
through actual physical constraints. Modeling these systems by spherical
particles interacting through vibrational potentials is obviously not
desirable as it may introduce spurious deformations and internal
oscillations.

As a result of the interest in constraints, much work has been devoted
in the literature
to develop practical, accurate, and efficient methods for introducing
constraints into the numerical methods to study the dynamics of
non-spherical particles.
Much of the pioneering work culminated in the SHAKE \cite{SHAKE} and
RATTLE \cite{RATTLE} algorithms. These methods consider the constraints
in a (point) particle ensemble as a collection of linearly independent
constraints on distances between pairs of particles. For any given time
step of a numerical temporal integrator of the equations of motion for each
particle, the constraints are enforced for the resulting positions through
the self-consistent solution of a set of nonlinear equations (one for each
constraint). This solution is usually obtained through an iterative approach
where computational efficiency has to be balanced with the desire for accuracy.
These approaches have been very successful and are still the core
in many applications. Some modifications have been introduced for specific
systems and geometries. For example, 
SETTLE \cite{SETTLE} provides a direct analytical
calculation of the three constraints necessary for rigid triangular objects
(developed for three-point classical water models),
and "Fast SHAKE" \cite{Fast_SHAKE} optimizes the iterative method for finding
the constraints necessary for modeling small rigid molecules.
Other suggestions include the EEM approach \cite{EEM,Yoneya}, which evaluates
the constraints non-iteratively at the beginning of a time step where all
positions are known. As a consequence of iterative methods and algorithmic
approximations, the updated positions do not exactly fulfill the
holonomic constraints, and corrective measures must be applied if
and when the cumulative discrepancy becomes unacceptable.

Most methods for evaluating constraints have been developed for
ensembles of linearly independent constraints. Several publications,
including \cite{SHAKE}, consider the linear geometry
as a separate case due to occurring singularities in the methods for
linearly independent constraints. A few direct methods for linear geometry
have been outlined (see, e.g., \cite{Ciccotti_82}), but they are
possible only because of perfect three-body symmetry.
We demonstrate a simple method for evaluating the necessary
Lagrange multipliers for constraining any number of different particles onto a
{\it linear} geometry. We show that the resulting method is efficient,
non-iterative, exact, and self-correcting.
\section{Equations of Motion with Constraints}
We consider the dynamics of $N$ particles constrained on a straight line
(see Figure 1). The equations of motion for the coordinate, $\bar{R}_j$, of
the $j$th particle is given by:
\begin{eqnarray}
m_j\ddot{\bar{R}}_j + \alpha_j\dot{\bar{R}}_j & = & \bar{{\cal F}}_{j} \; = \; 
\left\{\begin{array}{lcc}
\bar{f}_1 + \lambda_{1N}\bar{R}_{1N}-l_{1N}\displaystyle\sum_{k=2}^{N-1}\bar{\lambda}_{k1} & , & j=1 \\
\bar{f}_j + l_{1N}(\bar{\lambda}_{j1}+\bar{\lambda}_{jN}) & , & 1<j<N \\
\bar{f}_N - \lambda_{1N}\bar{R}_{1N}-l_{1N}\displaystyle\sum_{k=2}^{N-1}\bar{\lambda}_{kN} & , & j=N 
\end{array}\right. \; ,
\end{eqnarray}
where $m_j$ is the mass, friction is represented by $\alpha_j$, and $\bar{f}_j$
is an extra-molecular force on the $j$th particle.
Throughout the paper, overbar "$\bar{\;}$" and overdot "$\dot{\;}$" denote
a vector and a temporal derivative, respectively.
The two particles,
$j=1$ and $j=N$, define the direction and length of the object through the
scalar Lagrange multiplier $\lambda_{1N}$ and the vector
$\bar{R}_{1N}=\bar{R}_N-\bar{R}_1$. The length $l_{ji}$ is defined by
$l_{ji}=|\bar{R}_{i}-\bar{R}_{j}|$. All other particles, $1<j<N$, are
constrained to their relative positions through the vector multipliers
$\bar{\lambda}_{ji}$ for $i=1,N$, that connect each of those particles
exclusively to the defining particles of $j=1,N$. Notice that all the
forces of the constraints cancel for the ensemble.
At any given time the constraints are enforced by the geometric conditions:
\begin{eqnarray}
\left|\bar{R}_{1N}\right| & = & l_{1N} \\
\bar{R}_{j} & = & \frac{l_{jN}\bar{R}_{1}+l_{j1}\bar{R}_{N}}{l_{1N}} \; ,
\end{eqnarray}
where all particles $1<j<N$ lie between particles $1$
and $N$, such that $l_{j1}+l_{jN}=l_{1N}$. The number of geometrical
conditions given in (2) and (3) is $3(N-2)+1$ (three-dimensional space),
whereas the number of Lagrange multipliers to be determined for the solution
of (1) is $6(N-2)+1$. Since the constraints should not contribute net
torque to the object, we further have the conditions for $1<j<N$:
\begin{eqnarray}
\bar{R}_{1N}\times\left(l_{j1}\bar{\lambda}_{j1}-l_{jN}\bar{\lambda}_{jN}\right) & = & 0 \; .
\end{eqnarray}
By writing
\begin{eqnarray}
\bar{\lambda}_{ji} & = & \bar{\lambda}_{ji}^{\|}+\bar{\lambda}_{ji}^{\bot} \; ,
\end{eqnarray}
with $\bar{\lambda}_{ji}^{\bot}\cdot\bar{R}_{1N}=0$ and $\bar{\lambda}_{ji}^{\|}=\lambda_{ji}^{\|}\bar{R}_{1N}/|\bar{R}_{1N}|$, we can express (4)
in the form
\begin{eqnarray}
l_{j1}\bar{\lambda}_{j1}^\bot & = & l_{jN}\bar{\lambda}_{jN}^\bot \; ,
\end{eqnarray}
i.e., the constraint forces, responsible for maintaining particle $j$ at
its desired relative position, are balanced in their components orthogonal
to the direction of the object such that no net torque is contributed.

However, it is important to realize that this balance does not necessarily
apply to the longitudinal components, $\lambda_{j1}^{\|}$ and
$\lambda_{jN}^{\|}$. Since the object is rigid, we may therefore choose
{\it any} number, $\gamma_j$ ($1<j<N$), such that
\begin{eqnarray}
\gamma_j\bar{\lambda}_{j1}^\| & = & \bar{\lambda}_{jN}^\|
\end{eqnarray}
without any physical consequence. This ambiguity reflects the physical
fact that a rigid stick responds identically regardless of how
{\it longitudinal} forces are distributed along the object.
Inserting (6) and (7) into (1), we can thereby obtain the following
system of equations
\begin{eqnarray}
m_j\ddot{\bar{R}}_j + \alpha_j\dot{\bar{R}}_j & = & \left\{\begin{array}{lcc}
\bar{f}_1 + \lambda_{1N}\bar{R}_{1N}-l_{1N}\displaystyle\sum_{k=2}^{N-1}\left(\bar{\lambda}_{k1}^{\bot}+\bar{\lambda}_{k1}^{\|}\right) & , & j=1 \\
\bar{f}_j + l_{1N}\frac{l_{1N}}{l_{j1}}\bar{\lambda}_{j1}^\bot+l_{1N}(1+\gamma_{j})\bar{\lambda}_{j1}^\| & , & 1<j<N \\
\bar{f}_N - \lambda_{1N}\bar{R}_{1N}-l_{1N}\displaystyle\sum_{k=2}^{N-1}\left(\frac{l_{kN}}{l_{k1}}\bar{\lambda}_{k1}^{\bot}+\gamma_k\bar{\lambda}_{k1}^{\|}\right) & , & j=N
\end{array}\right. \; ,
\end{eqnarray}
with $3(N-2)+1$ Lagrange multipliers, determined by the geometry of (2)
and (3) (notice that the vector
$\bar{\lambda}_{k1}=\bar{\lambda}_{k1}^\bot+\bar{\lambda}_{k1}^\|$
represents 3 scalar multipliers, one for the longitudinal direction
$\bar{\lambda}_{k1}^\|$ and two for orthogonal directions
$\bar{\lambda}_{k1}^\bot$).
These equations can now be introduced into a variety of numerical
integrators, determining the Lagrange multipliers for enforcing the
constraints, and exploiting the free Gauge parameters, $\gamma_j$,
for computational optimization.
\section{Using the Velocity Explicit Verlet Method}
In the following we illustrate the approach through the commonly
used \cite{overview_2,RATTLE} velocity Explicit Verlet algorithm
(shown here including linear
damping) which approximates the solution
to (1) using a non-zero time step of $dt$ connecting time $t_n=ndt$ to
$t_{n+1}=(n+1)dt$,
\begin{eqnarray}
\bar{R}^{n+1}_j & = & \bar{R}^{n}_j + \left(1-\frac{\alpha_j dt}{2}\right)dt\bar{V}^n_j+\frac{dt^2}{2m_j}\bar{{\cal F}}_j^n \\
\bar{V}^{n+1}_j & = & \frac{1-\frac{\alpha_j dt}{2}}{1+\frac{\alpha_j dt}{2}} \bar{V}^{n}_j + \frac{1}{1+\frac{\alpha_j dt}{2}}\frac{dt}{2m_j}\left(\bar{{\cal F}}_j^n+\bar{{\cal F}}_j^{n+1}\right) \; ,
\end{eqnarray}
where we use the notation,
$\bar{X}^n_j=\bar{X}_j(t_n)=\bar{X}_j(n\cdot dt)$, to describe
the integer $n$ time step, and where $\bar{V}_j=\dot{\bar{R}}_j$.
\subsection{Determining the Lagrange Multipliers $\bar{\lambda}_{ji}$}
Inserting (8) into (9) yields the constrained discrete-time
equations 
\begin{eqnarray}
\bar{R}^{n+1}_j & = & \tilde{R}_j^{n+1} + {\cal A}_j \times \left\{\begin{array}{rcc}
\lambda_{1N}\bar{R}_{1N}^n-l_{1N}\displaystyle\sum_{k=2}^{N-1}
\left(\bar{\lambda}_{k1}^{\bot}+\bar{\lambda}_{k1}^{\|}\right) & , & j=1 \\
l_{1N}\frac{l_{1N}}{l_{j1}}\bar{\lambda}_{j1}^\bot+l_{1N}(1+\gamma_{j})\bar{\lambda}_{j1}^\| & , & 1<j<N \\
-\lambda_{1N}\bar{R}_{1N}^n-l_{1N}\displaystyle\sum_{k=2}^{N-1}\left(\frac{l_{kN}}{l_{k1}}\bar{\lambda}_{k1}^{\bot}+\gamma_k\bar{\lambda}_{k1}^{\|}\right) & , & j=N
\end{array}\right. \; ,
\end{eqnarray}
with ${\cal A}_j=\frac{dt^2}{2m_j}$ and, following the notation of
\cite{overview_2,RATTLE},
$\tilde{R}_j^{n+1}$ is the updated (vector) position of particle $j$, had
there been no
constraints. The longitudinal and orthogonal components of $\bar{\lambda}_{ji}$
are here relative to $\bar{R}_{1N}^n$. The Lagrange multipliers are now
determined such that the constraints are satisfied at time $t_{n+1}$; i.e.,
under the conditions
\begin{eqnarray}
\left|\bar{R}_{1N}^{n+1}\right|^2 & = & l_{1N}^2 \\
1<j<N: \; \; \; \; \bar{R}_{j}^{n+1} & = & \frac{l_{jN}\bar{R}_1^{n+1}+l_{j1}\bar{R}_{N}^{n+1}}{l_{1N}} \; .
\end{eqnarray}

Inserting (11) into (12) provides us with one
equation for the determination of $\lambda_{1N}$,
\begin{eqnarray}
l_{1N}^2 & = & \Big|\overbrace{\tilde{R}_{1N}^{n+1}
+l_{1N}\sum_{k=2}^{N-1}\left({\cal A}_1-{\cal A}_2\frac{l_{k1}}{l_{k2}}\right)
\bar{\lambda}_{k1}^\bot}^{\bar{\rho}}+\underbrace{l_{1N}\sum_{k=2}^{N-1}\left({\cal A}_1-{\cal A}_N\gamma_k\right)\bar{\lambda}_{k1}^\|}_{=0 \; \; {\rm for} \; \; \gamma_k={\cal A}_1/{\cal A}_N}-({\cal A}_1+{\cal A}_N)\bar{R}_{1N}^n\lambda_{1N}\Big|^2 \; , \nonumber \\
\end{eqnarray}
whereas inserting (11) into (13) provides us with $3(N-2)$ equations
for the determination of $\bar{\lambda}_{j1}$,
\begin{eqnarray}
-{\cal A}_j\frac{l_{1N}^2}{l_{jN}}\bar{\lambda}_{j1}^\bot-\sum_{k=2}^{N-1}\left(l_{jN}{\cal A}_1+l_{j1}\frac{l_{k1}}{l_{kN}}{\cal A}_N\right)\bar{\lambda}_{k1}^\bot & = & \Delta\tilde{R}_j^\bot \\
-{\cal A}_j l_{1N}(1+\gamma_j)\bar{\lambda}_{j1}^\|-\sum_{k=2}^{N-1}(l_{1N}{\cal A}_1-\underbrace{l_{j1}\left({\cal A}_1-\gamma_k{\cal A}_N\right)}_{=0 \; \; {\rm for} \; \; \gamma_k={\cal A}_1/{\cal A}_N})\bar{\lambda}_{k1}^\| & = & \Delta\tilde{R}_j^\|-\frac{\bar{R}_{1N}^n}{l_{1N}}\left({\cal A}_1 l_{jN}-{\cal A}_N l_{j1}\right)\lambda_{1N} \; , \nonumber \\
\end{eqnarray}
where
\begin{eqnarray}
\Delta\tilde{R}_j & = & \Delta\tilde{R}_j^\bot + \Delta\tilde{R}_j^\| \frac{\bar{R}_{1N}^n}{\left|\bar{R}_{1N}^n\right|} \; = \; \tilde{R}_j^{n+1}-\frac{l_{jN}\tilde{R}_1^{n+1}+l_{j1}\tilde{R}_{N}^{n+1}}{l_{1N}} \; .
\end{eqnarray}
Solving equations (14)-(16) will provide the Lagrange multipliers. However,
since (14) is a nonlinear equation, and since (14)-(16) are coupled,
the direct solution to the equations will require an iterative approach
to the solution of the $3(N-2)+1$ equations. A significant computational
consequence of separating the orthogonal and longitudinal components of the
Lagrange multiplier $\bar{\lambda}_{j1}$ is that the "free" parameters
$\gamma_j$ (for $1<j<N$) introduced in (7) can now be chosen to
optimize efficiency. Specifically, choosing
{\underline{$\gamma_j={\cal A}_1/{\cal A}_N=m_N/m_1$}} leads to the following
{\it efficient}, {\it non-iterative}, {\it exact}, and {\it self-correcting}
procedure for evaluating the necessary multipliers:
\begin{description}
\item[Step 1:] Solve the two independent
sets of $N-2$ linear equations (15) to obtain $\bar{\lambda}_{j1}^\bot$.
\item[Step 2:] Solve (14). With $\bar{\lambda}_{j1}^\bot$ evaluated, then
$\bar{\rho}$ is known.
Given $\gamma_j={\cal A}_1/{\cal A}_N$, this equation becomes a simple
second-order polynomial in $\lambda_{1N}$ where the root closer to zero
is the relevant one.
\item[Step 3:] With the completion of steps 1 and 2, $\lambda_{j1}^\|$
can be found from (16) (with $\gamma_j={\cal A}_1/{\cal A}_N$)
as the solution to the set of $N-2$ linear equations.
\end{description}
The facts that the method outlined above is non-iterative and exact are
self-evident. Given that the involved matrices are rather small
($(N-2)\times(N-2)$), the method is also computationally efficient.
Finally, it can be seen that regardless of deviations from the desired
constrained geometry at time $n\cdot dt$, the geometric conditions for
evaluating the Lagrange multipliers are enforced at time $(n+1)\cdot dt$
(see (12) and (13)). Thus, the method is self-correcting when making the
important distinction between, e.g., a {\it desired} length $l_{j1}$ and
an {\it actual} length $|\bar{R}_{j1}^n|$, which may not be equal to $l_{j1}$
due to computational precision errors.

The matrices corresponding to the $N-2$ equations in (15) and (16) are
generally very well conditioned and are not subject to numerical instabilities.
Specifically, for a system of $N$ equal masses
the matrix described in (16) has condition number \cite{condition}
$k_2^\|=\frac{1}{2}N$ (for $N>3$).
The matrix (15) describing the orthogonal components of the Lagrange multipliers
is less transparent. However, for a system of evenly spaced equal masses,
we show in Figure 2 how the condition number $k_2^\bot$ depends on $N$.
It is clear from these data that the matrices are easily handled numerically
for any reasonable number of constrained particles. It is important to realize
that the matrix (15) may become less well conditioned if ${\cal A}_N/l_{jN}$
is extremely large for any $0<j<N$. However, (11) and (15) have been expressed
in terms of $\bar{\lambda}_{j1}$, through the relationships (6) and
(7). Thus, we could equally well have chosen the sought after Lagrange
multipliers to be $\bar{\lambda}_{jN}$, in which case equations
equivalent to (15)-(17) can be produced to avoid a possible
ill conditioned matrix for large ${\cal A}_N/l_{jN}$.
It is also important to realize that these matrices are not changing during the
course of a simulation, and that, e.g., the same $LU$ factorization can be
effectively used for all time steps.
The solution to (14) is limited mainly by $\bar{\rho}$, as it can be seen
that the polynomial may have only complex solutions when $\bar{\rho}$ is near
orthogonal to $\bar{R}_{1N}^n$. For reasonable time steps, when particle
positions only change by a fraction of particle size, $\bar{\rho}$ is always
near parallel to $\bar{R}_{1N}^n$, and (14) does therefore not pose any
unpredictable numerical problems.

The completion of the time step in (11) can then be accomplished for all
participating particles in an ensemble. With these new coordinates at time
$t_{n+1}$, the resulting forces $\bar{f}_j^{n+1}$ can be evaluated.
\subsection{Updating the Velocities}
As pointed out in \cite{RATTLE}, straightforward application of
(10) will not in general satisfy the necessary constraints on the
velocity complements to (12) and (13):
\begin{eqnarray}
\bar{V}_{1N}^{n+1}\cdot\bar{R}_{1N}^{n+1} & = & 0 \\
1<j<N: \; \; \; \; \bar{V}_{j}^{n+1} & = & \frac{l_{jN}\bar{V}_1^{n+1}+l_{j1}\bar{V}_{N}^{n+1}}{l_{1N}} \; .
\end{eqnarray}
The necessary corrections can be introduced by the following modification,
\begin{eqnarray}
\bar{{\cal F}}_j^{n+1} & = & \bar{f}_j^{n+1} + {\cal B}_j \times
\left\{\begin{array}{rcc}
\sigma_{1N}\bar{R}_{1N}^{n+1}-l_{1N}\displaystyle\sum_{k=2}^{N-1}\bar{\sigma}_{k1} & , & j=1 \\
l_{1N}\left(\bar{\sigma}_{j1}+\bar{\sigma}_{jN}\right) & , & 1<j<N \\
-\sigma_{1N}\bar{R}_{1N}^{n+1}-l_{1N}\displaystyle\sum_{k=2}^{N-1}\bar{\sigma}_{k1} & , & j=N 
\end{array}\right. \; ,
\end{eqnarray}
where ${\cal B}_j=dt/(2m_j+\alpha_j dt)$. As for the Lagrange multipliers
$\bar{\lambda}_{ji}$ the introduction of $\bar{\sigma}_{ji}$ should conserve
angular momentum, i.e.,
\begin{eqnarray}
  l_{j1}\bar{\sigma}_{j1}^\bot & = & l_{jN}\bar{\sigma}_{jN}^\bot \\
\delta_j\bar{\sigma}_{j1}^\|   & = &       \bar{\sigma}_{jN}^\|  \; ,
\end{eqnarray}
where $\bar{\sigma}_{ji}=\bar{\sigma}_{ji}^\bot+\bar{\sigma}_{ji}^\|$ with
$\bar{\sigma}_{ji}^\bot\cdot\bar{R}_{1N}^{n+1}=0$ and 
$\bar{\sigma}_{ji}^\| \| \bar{R}_{1N}^{n+1}$. Analogous to the physically
arbitrary coefficient $\gamma_j$, here we have coefficients $\delta_j$
that can be chosen to be any number. With this observation, (20) becomes
the velocity complement to (11):
\begin{eqnarray}
\bar{V}^{n+1}_j & = & \tilde{V}_j^{n+1} + {\cal B}_j \times \left\{\begin{array}{rcc}
\sigma_{1N}\bar{R}_{1N}^{n+1}-l_{1N}\displaystyle\sum_{k=2}^{N-1}
\left(\bar{\sigma}_{k1}^{\bot}+\bar{\sigma}_{k1}^{\|}\right) & , & j=1 \\
l_{1N}\frac{l_{1N}}{l_{j1}}\bar{\sigma}_{j1}^\bot+l_{1N}(1+\delta_{j})\bar{\sigma}_{j1}^\| & , & 1<j<N \\
-\sigma_{1N}\bar{R}_{1N}^{n+1}-l_{1N}\displaystyle\sum_{k=2}^{N-1}\left(\frac{l_{kN}}{l_{k1}}\bar{\sigma}_{k1}^{\bot}+\delta_k\bar{\sigma}_{k1}^{\|}\right) & , & j=N
\end{array}\right. \; .
\end{eqnarray}
The (vector) velocities without the corrective multipliers
$\bar{\sigma}$ are denoted $\tilde{V}_j^{n+1}$.
The velocity equations are now determined by inserting (23) into
(18) and (19), yielding the following set of $3(N-2)+1$ linear
equations,
\begin{eqnarray}
\Delta\tilde{V}_j^\bot & = & -{\cal B}_j\frac{l_{1N}^2}{l_{jN}}\bar{\sigma}_{j1}^\bot-\sum_{k=2}^{N-1}\left(l_{jN}{\cal B}_1+l_{j1}\frac{l_{k1}}{l_{kN}}{\cal B}_N\right)\bar{\sigma}_{k1}^\bot \\
\sigma_{1N} & = & \frac{\bar{R}_{1N}^{n+1}\cdot\tilde{V}_{1N}^{n+1}+\overbrace{l_{1N}\displaystyle\sum_{k=2}^{N-1}({\cal B}_1-\delta_k{\cal B}_N)\bar{\sigma}_{k1}}^{=0 \; \; {\rm for} \; \; \delta_k={\cal B}_1/{\cal B}_N}}{\left|\bar{R}_{1N}^{n+1}\right|^2 ({\cal B}_1+{\cal B}_N)} \\
\Delta\tilde{V}_j^\| - \frac{\bar{R}_{1N}^{n+1}}{l_{1N}}\left({\cal B}_1 l_{jN}-{\cal B}_N l_{j1}\right)\sigma_{1N} & = & -{\cal B}_j l_{1N}(1+\delta_j)\bar{\sigma}_{j1}^\|-\sum_{k=2}^{N-1}(l_{1N}{\cal B}_1-\underbrace{l_{j1}\left({\cal B}_1-\delta_k{\cal B}_N\right)}_{=0 \; \; {\rm for} \; \; \delta_k={\cal B}_1/{\cal B}_N})\bar{\sigma}_{k1}^\| \; , \nonumber \\
\end{eqnarray}
where
\begin{eqnarray}
\Delta\tilde{V}_j & = & \Delta\tilde{V}_j^\bot + \Delta\tilde{V}_j^\| \frac{\bar{R}_{1N}^{n+1}}{\left|\bar{R}_{1N}^{n+1}\right|} \; = \; \Delta\tilde{V}_j^{n+1}-\frac{l_{jN}\Delta\tilde{V}_1^{n+1}+l_{j1}\Delta\tilde{V}_{N}^{n+1}}{l_{1N}} \; .
\end{eqnarray}

The solution to the set of linear equations (24)-(26) can be significantly
assisted by choosing {\underline{$\delta_k={\cal B}_1/{\cal B}_N$}.
This choice will allow (25) to become
independent of the other equations, and all corrective multipliers $\sigma$
can be obtained by solving three sets of $N-2$ (plus one) linear
equations in the listed
order. With this solution, the velocities $\bar{V}_j^{n+1}$ can be completed
through (23), and the time step is then accomplished. Before initiating
another time step, it is important to include the corrective multipliers
into the evaluated force, i.e., $\bar{f}_j^{n+1}=\bar{\cal F}_j^{n+1}$,
before evaluating $\tilde{R}_j^{n+2}$.

Notice that for small $dt$, the matrices given by (24) and (26) are
nearly identical to the ones described by (15) and (16), respectively.
Thus, we can assume that (24) and (26) are subject to very similar
conditioning as discussed in the previous section,
when solving the systems numerically.
\section{Conclusion}
We have presented an efficient algorithm for conducting dynamical
simulations of particles constrained in a linear geometry. Following
other presentations on geometrical constraints in particle simulations,
we introduce Lagrange multipliers to the equations of motion for each particle,
and determine the multipliers by the desired constraints as well as basic
conditions of conservation of momentum and angular momentum. The resulting
equations for determining the Lagrange multipliers are, in general, a set of
coupled equations with some nonlinear component, necessitating an
iterative approach to obtain an approximate solution. By taking advantage
of the physical ambiguity left in the distribution of the constraining forces
in the longitudinal direction of the ensemble, we are able to determine
a particular distribution that decouples the nonlinearity of the determining
equations into a second-order polynomial of a single variable, namely,
one of the Lagrange multipliers in the longitudinal direction. The rest of
the equations are linear and can be solved as three independent sets of
$N-2$ equations with $N-2$ unknowns, where $N$ is the number of mutually
constrained particles.
Thus, the algorithm is exact (within the accuracy of floating-point arithmetic),
efficient, and non-iterative. We have illustrated the approach through an
implementation with the widely used velocity explicit
Verlet algorithm for temporal integration
of second-order differential equations. The presented formulation of the
approach has the added benefit of being self-correcting, i.e., any
spurious deviations in the constrained geometry arising from, e.g.,
computational precision error, is self-corrected, and the algorithm
does not accumulate or propagate error in the constraints. It is
straightforward to apply the method to other numerical methods for ordinary
differential equations. For example, a recent comprehensive study on
self-assembled monolayers \cite{Jensen_03} applied the above-outlined
technique to model rigid and linear alkanethiol molecules as overdamped
stochastically driven objects.
\section{Acknowledgments}
This research was supported by funds from the California Department of 
Transportation through its Partnered Pavement Research Center project at 
the University of California, Davis and Berkeley campuses.
The work was also supported by NSF Biophotonics Science and
Technology Center (UC~Davis), and by the CARE program of
Los Alamos National Laboratory. We are grateful to Prof.~Alan Laub
for carefully reading the manuscript and providing several useful suggestions.

\begin{figure}
\epsfxsize=5.75in
\epsfbox{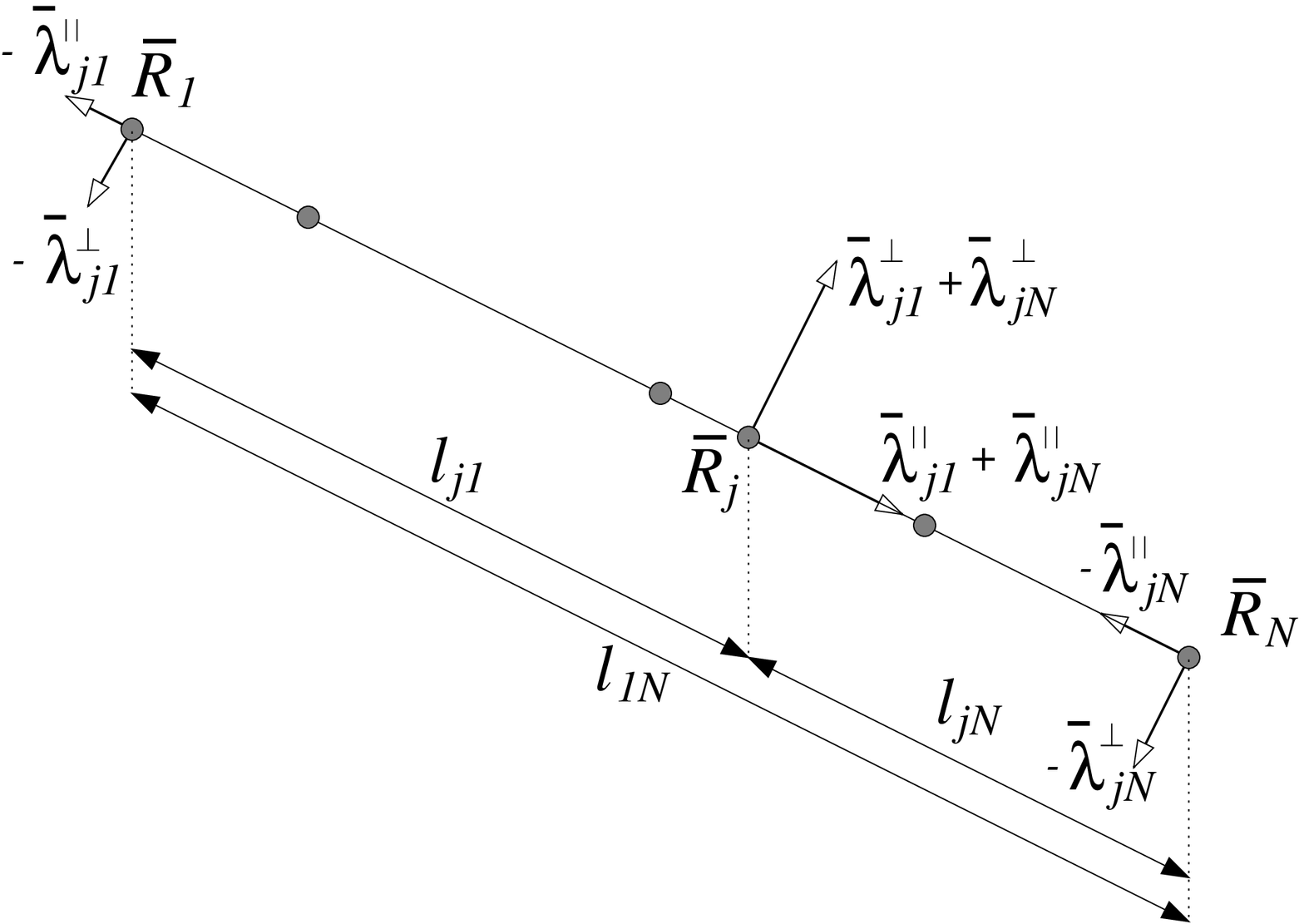}
\caption{Sketch of the system under consideration. $N$ particles are
constrained to the line, defined by $\bar{R}_N-\bar{R}_1$, through the
Lagrange multipliers $\bar{\lambda}_{ji}=\bar{\lambda}_{ji}^\bot+\bar{\lambda}_{ji}^\|$, where $1<j<N$ and $i=1,N$.}
\label{fig:fig1}
\end{figure}

\begin{figure}
\epsfxsize=5.75in
\epsfbox{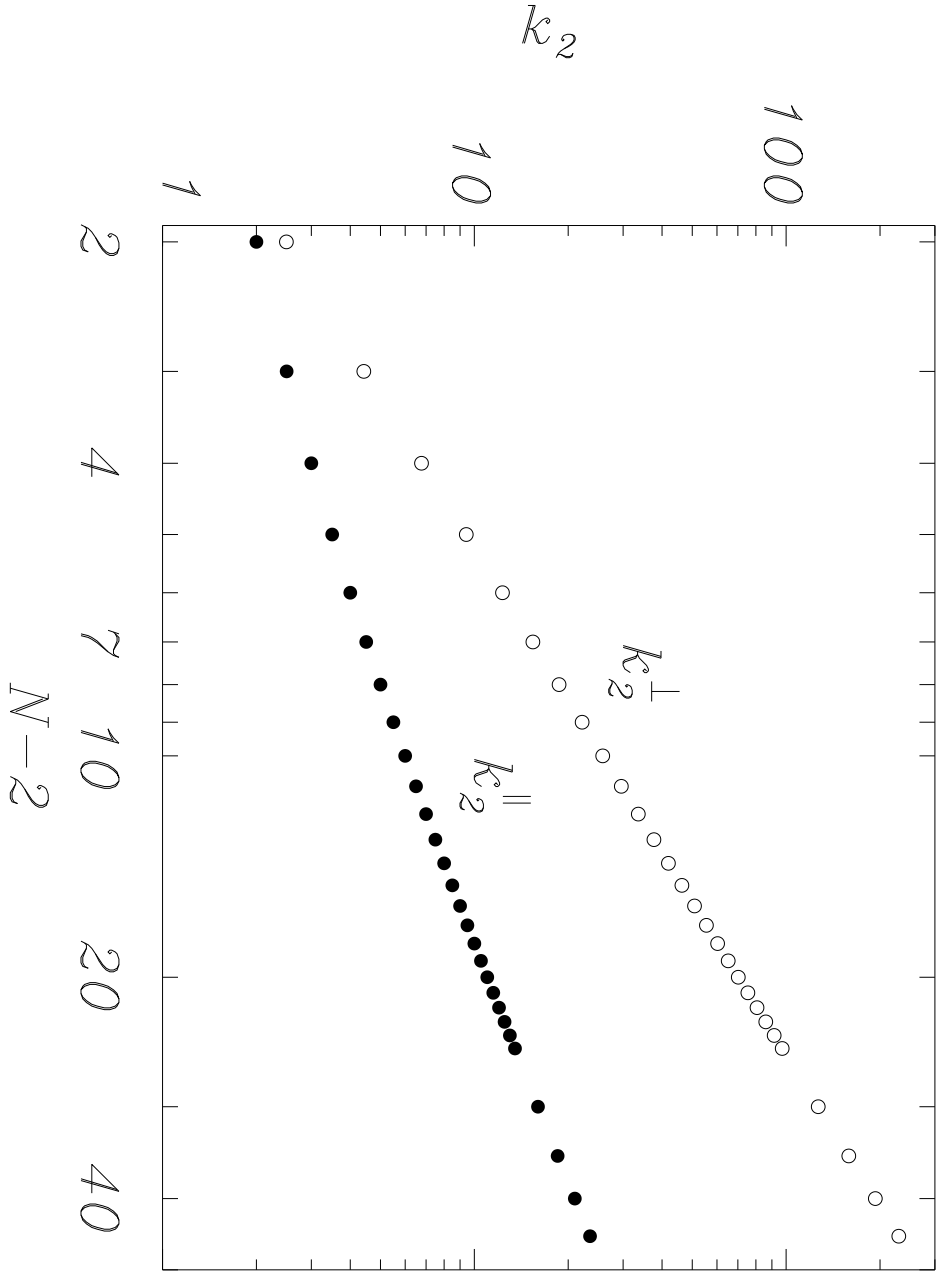}
\caption{Condition numbers, $k_2^\bot$ and $k_2^\|$,
for the matrices described by equation (15) ($\circ$) and equation (16)
($\bullet$) for $\gamma_k={\cal A}_1/{\cal A}_N$ as a function of the
matrix size $(N-2)\times(N-2)$, if the
$N$ particles have equal mass and are evenly spaced along $\bar{R}_{1N}$.
The condition numbers are $k_2^\|=\frac{1}{2}N$ and
$k_2^\bot\approx(N-2)^{1.4}$.}
\label{fig:fig2}
\end{figure}
\end{document}